\overfullrule=0pt
%
%
%
%
%
\def\unredoffs{} 

%
%
%
%
%
\newbox\leftpage \newdimen\fullhsize \newdimen\hstitle \newdimen\hsbody
\tolerance=1000\hfuzz=2pt
\catcode`\@=11 
\def\bigans{b }
%
\magnification=1200\unredoffs\baselineskip=16pt plus 2pt minus 1pt
\hsbody=\hsize \hstitle=\hsize 
%
%
%
\newcount\yearltd\yearltd=\year\advance\yearltd by -1900

\def\Title#1#2{\nopagenumbers\abstractfont\hsize=\hstitle\rightline{#1}%
\vskip 1in\centerline{\titlefont #2}\abstractfont\vskip .5in\pageno=0}
\def\Date#1{\vfill\leftline{#1}\tenpoint\supereject\global\hsize=\hsbody%
\footline={\hss\tenrm\folio\hss}}
%

\def\draftmode{\message{ DRAFTMODE }\def\draftdate{{\rm preliminary draft:
\number\month/\number\day/\number\yearltd\ \ \hourmin}}%
\headline={\hfil\draftdate}\writelabels\baselineskip=20pt plus 2pt minus 2pt
 {\count255=\time\divide\count255 by 60 \xdef\hourmin{\number\count255}
  \multiply\count255 by-60\advance\count255 by\time
  \xdef\hourmin{\hourmin:\ifnum\count255<10 0\fi\the\count255}}}
\def\nolabels{\def\wrlabeL##1{}\def\eqlabeL##1{}\def\reflabeL##1{}}
\def\writelabels{\def\wrlabeL##1{\leavevmode\vadjust{\rlap{\smash%
{\line{{\escapechar=` \hfill\rlap{\sevenrm\hskip.03in\string##1}}}}}}}%
\def\eqlabeL##1{{\escapechar-1\rlap{\sevenrm\hskip.05in\string##1}}}%
\def\reflabeL##1{\noexpand\llap{\noexpand\sevenrm\string\string\string##1}}}
\nolabels
%
\global\newcount\secno \global\secno=0
\global\newcount\meqno \global\meqno=1
\def\newsec#1{\global\advance\secno by1\message{(\the\secno. #1)}
\global\subsecno=0\eqnres@t\noindent{\bf\the\secno. #1}
\writetoca{{\secsym} {#1}}\par\nobreak\medskip\nobreak}
\def\eqnres@t{\xdef\secsym{\the\secno.}\global\meqno=1\bigbreak\bigskip}
\def\sequentialequations{\def\eqnres@t{\bigbreak}}\xdef\secsym{}
\global\newcount\subsecno \global\subsecno=0
\def\subsec#1{\global\advance\subsecno by1\message{(\secsym\the\subsecno. #1)}
\ifnum\lastpenalty>9000\else\bigbreak\fi
\noindent{\it\secsym\the\subsecno. #1}\writetoca{\string\quad 
{\secsym\the\subsecno.} {#1}}\par\nobreak\medskip\nobreak}
\def\appendix#1#2{\global\meqno=1\global\subsecno=0\xdef\secsym{\hbox{#1.}}
\bigbreak\bigskip\noindent{\bf Appendix #1. #2}\message{(#1. #2)}
\writetoca{Appendix {#1.} {#2}}\par\nobreak\medskip\nobreak}
%
%
\def\eqnn#1{\xdef #1{(\secsym\the\meqno)}\writedef{#1\leftbracket#1}%
\global\advance\meqno by1\wrlabeL#1}
\def\eqna#1{\xdef #1##1{\hbox{$(\secsym\the\meqno##1)$}}
\writedef{#1\numbersign1\leftbracket#1{\numbersign1}}%
\global\advance\meqno by1\wrlabeL{#1$\{\}$}}
\def\eqn#1#2{\xdef #1{(\secsym\the\meqno)}\writedef{#1\leftbracket#1}%
\global\advance\meqno by1$$#2\eqno#1\eqlabeL#1$$}
%
\newskip\footskip\footskip14pt plus 1pt minus 1pt 
\def\footnotefont{\ninepoint}\def\f@t#1{\footnotefont #1\@foot}
\def\f@@t{\baselineskip\footskip\bgroup\footnotefont\aftergroup\@foot\let\next}
\setbox\strutbox=\hbox{\vrule height9.5pt depth4.5pt width0pt}
\global\newcount\ftno \global\ftno=0
\def\foot{\global\advance\ftno by1\footnote{$^{\the\ftno}$}}
%
\newwrite\ftfile   
\def\footend{\def\foot{\global\advance\ftno by1\chardef\wfile=\ftfile
$^{\the\ftno}$\ifnum\ftno=1\immediate\openout\ftfile=foots.tmp\fi%
\immediate\write\ftfile{\noexpand\smallskip%
\noexpand\item{f\the\ftno:\ }\pctsign}\findarg}%
\def\footatend{\vfill\eject\immediate\closeout\ftfile{\parindent=20pt
\centerline{\bf Footnotes}\nobreak\bigskip\input foots.tmp }}}
\def\footatend{}
%
%
\global\newcount\refno \global\refno=1
\newwrite\rfile
\def\ref{[\the\refno]\nref}
\def\nref#1{\xdef#1{[\the\refno]}\writedef{#1\leftbracket#1}%
\ifnum\refno=1\immediate\openout\rfile=refs.tmp\fi
\global\advance\refno by1\chardef\wfile=\rfile\immediate
\write\rfile{\noexpand\item{#1\ }\reflabeL{#1\hskip.31in}\pctsign}\findarg}
\def\findarg#1#{\begingroup\obeylines\newlinechar=`\^^M\pass@rg}
{\obeylines\gdef\pass@rg#1{\writ@line\relax #1^^M\hbox{}^^M}%
\gdef\writ@line#1^^M{\expandafter\toks0\expandafter{\striprel@x #1}%
\edef\next{\the\toks0}\ifx\next\em@rk\let\next=\endgroup\else\ifx\next\empty%
\else\immediate\write\wfile{\the\toks0}\fi\let\next=\writ@line\fi\next\relax}}
\def\striprel@x#1{} \def\em@rk{\hbox{}} 
\def\lref{\begingroup\obeylines\lr@f}
\def\lr@f#1#2{\gdef#1{\ref#1{#2}}\endgroup\unskip}
\def\semi{;\hfil\break}
\def\addref#1{\immediate\write\rfile{\noexpand\item{}#1}} 
\def\startrefs#1{\immediate\openout\rfile=refs.tmp\refno=#1}
\def\xref{\expandafter\xr@f}\def\xr@f[#1]{#1}
\def\refs#1{\count255=1[\r@fs #1{\hbox{}}]}
\def\r@fs#1{\ifx\und@fined#1\message{reflabel \string#1 is undefined.}%
\nref#1{need to supply reference \string#1.}\fi%
\vphantom{\hphantom{#1}}\edef\next{#1}\ifx\next\em@rk\def\next{}%
\else\ifx\next#1\ifodd\count255\relax\xref#1\count255=0\fi%
\else#1\count255=1\fi\let\next=\r@fs\fi\next}
%

%
\newwrite\ffile\global\newcount\figno \global\figno=1
\def\fig{fig.~\the\figno\nfig}
\def\nfig#1{\xdef#1{fig.~\the\figno}%
\writedef{#1\leftbracket fig.\noexpand~\the\figno}%
\ifnum\figno=1\immediate\openout\ffile=figs.tmp\fi\chardef\wfile=\ffile%
\immediate\write\ffile{\noexpand\medskip\noexpand\item{Fig.\ \the\figno. }
\reflabeL{#1\hskip.55in}\pctsign}\global\advance\figno by1\findarg}
\def\xfig{\expandafter\xf@g}\def\xf@g fig.\penalty\@M\ {}
\def\figs#1{figs.~\f@gs #1{\hbox{}}}
\def\f@gs#1{\edef\next{#1}\ifx\next\em@rk\def\next{}\else
\ifx\next#1\xfig #1\else#1\fi\let\next=\f@gs\fi\next}
\newwrite\lfile
{\escapechar-1\xdef\pctsign{\string\%}\xdef\leftbracket{\string\{}
\xdef\rightbracket{\string\}}\xdef\numbersign{\string\#}}

\def\writestop{\def\writestoppt{\immediate\write\lfile{\string\pageno%
\the\pageno\string\startrefs\leftbracket\the\refno\rightbracket%
\string\def\string\secsym\leftbracket\secsym\rightbracket%
\string\secno\the\secno\string\meqno\the\meqno}\immediate\closeout\lfile}}
\def\writestoppt{}\def\writedef#1{}
\def\seclab#1{\xdef #1{\the\secno}\writedef{#1\leftbracket#1}\wrlabeL{#1=#1}}
\def\subseclab#1{\xdef #1{\secsym\the\subsecno}%
\writedef{#1\leftbracket#1}\wrlabeL{#1=#1}}
\newwrite\tfile \def\writetoca#1{}
\def\leaderfill{\leaders\hbox to 1em{\hss.\hss}\hfill}
\def\writetoc{\immediate\openout\tfile=toc.tmp 
   \def\writetoca##1{{\edef\next{\write\tfile{\noindent ##1 
   \string\leaderfill {\noexpand\number\pageno} \par}}\next}}}

%
\catcode`\@=12 
%
\edef\tfontsize{\ifx\answ\bigans scaled\magstep3\else scaled\magstep4\fi}
\font\titlerm=cmr10 \tfontsize \font\titlerms=cmr7 \tfontsize
\font\titlermss=cmr5 \tfontsize \font\titlei=cmmi10 \tfontsize
\font\titleis=cmmi7 \tfontsize \font\titleiss=cmmi5 \tfontsize
\font\titlesy=cmsy10 \tfontsize \font\titlesys=cmsy7 \tfontsize
\font\titlesyss=cmsy5 \tfontsize \font\titleit=cmti10 \tfontsize
\skewchar\titlei='177 \skewchar\titleis='177 \skewchar\titleiss='177
\skewchar\titlesy='60 \skewchar\titlesys='60 \skewchar\titlesyss='60
\def\titlefont{\def\rm{\fam0\titlerm}
\textfont0=\titlerm \scriptfont0=\titlerms \scriptscriptfont0=\titlermss
\textfont1=\titlei \scriptfont1=\titleis \scriptscriptfont1=\titleiss
\textfont2=\titlesy \scriptfont2=\titlesys \scriptscriptfont2=\titlesyss
\textfont\itfam=\titleit \def\it{\fam\itfam\titleit}\rm}
 \ifx\answ\bigans\else scaled\magstep1\fi
\ifx\answ\bigans\def\abstractfont{\tenpoint}\else
\font\abssl=cmsl10 scaled \magstep1
\font\absrm=cmr10 scaled\magstep1 \font\absrms=cmr7 scaled\magstep1
\font\absrmss=cmr5 scaled\magstep1 \font\absi=cmmi10 scaled\magstep1
\font\absis=cmmi7 scaled\magstep1 \font\absiss=cmmi5 scaled\magstep1
\font\abssy=cmsy10 scaled\magstep1 \font\abssys=cmsy7 scaled\magstep1
\font\abssyss=cmsy5 scaled\magstep1 \font\absbf=cmbx10 scaled\magstep1
\skewchar\absi='177 \skewchar\absis='177 \skewchar\absiss='177
\skewchar\abssy='60 \skewchar\abssys='60 \skewchar\abssyss='60
\def\abstractfont{\def\rm{\fam0\absrm}
\textfont0=\absrm \scriptfont0=\absrms \scriptscriptfont0=\absrmss
\textfont1=\absi \scriptfont1=\absis \scriptscriptfont1=\absiss
\textfont2=\abssy \scriptfont2=\abssys \scriptscriptfont2=\abssyss
\textfont\itfam=\bigit \def\it{\fam\itfam\bigit}\def\footnotefont{\tenpoint}%
\textfont\slfam=\abssl \def\sl{\fam\slfam\abssl}%
\textfont\bffam=\absbf \def\bf{\fam\bffam\absbf}\rm}\fi
\def\tenpoint{\def\rm{\fam0\tenrm}
\textfont0=\tenrm \scriptfont0=\sevenrm \scriptscriptfont0=\fiverm
\textfont1=\teni  \scriptfont1=\seveni  \scriptscriptfont1=\fivei
\textfont2=\tensy \scriptfont2=\sevensy \scriptscriptfont2=\fivesy
\textfont\itfam=\tenit \def\it{\fam\itfam\tenit}\def\footnotefont{\ninepoint}%
\textfont\bffam=\tenbf \def\bf{\fam\bffam\tenbf}\def\sl{\fam\slfam\tensl}\rm}
\font\ninerm=cmr9 \font\sixrm=cmr6 \font\ninei=cmmi9 \font\sixi=cmmi6 
\font\ninesy=cmsy9 \font\sixsy=cmsy6 \font\ninebf=cmbx9 
\font\nineit=cmti9 \font\ninesl=cmsl9 \skewchar\ninei='177
\skewchar\sixi='177 \skewchar\ninesy='60 \skewchar\sixsy='60 
\def\ninepoint{\def\rm{\fam0\ninerm}
\textfont0=\ninerm \scriptfont0=\sixrm \scriptscriptfont0=\fiverm
\textfont1=\ninei \scriptfont1=\sixi \scriptscriptfont1=\fivei
\textfont2=\ninesy \scriptfont2=\sixsy \scriptscriptfont2=\fivesy
\textfont\itfam=\ninei \def\it{\fam\itfam\nineit}\def\sl{\fam\slfam\ninesl}%
\textfont\bffam=\ninebf \def\bf{\fam\bffam\ninebf}\rm} 
%
%

\hyphenation{anom-aly anom-alies coun-ter-term coun-ter-terms}
\def\inv{^{\raise.15ex\hbox{${\scriptscriptstyle -}$}\kern-.05em 1}}

\def\Dsl{\,\raise.15ex\hbox{/}\mkern-13.5mu D} 
\def\dsl{\raise.15ex\hbox{/}\kern-.57em\partial}

\font\bigit=cmti10 scaled \magstep1
\def\lspace{\ifx\answ\bigans{}\else\qquad\fi}
\def\lbspace{\ifx\answ\bigans{}\else\hskip-.2in\fi} 
\def\boxeqn#1{\vcenter{\vbox{\hrule\hbox{\vrule\kern3pt\vbox{\kern3pt
	\hbox{${\displaystyle #1}$}\kern3pt}\kern3pt\vrule}\hrule}}}
\def\mbox#1#2{\vcenter{\hrule \hbox{\vrule height#2in
		\kern#1in \vrule} \hrule}}  
%

\def\e#1{{\rm e}^{^{\textstyle#1}}}

\def\darr#1{\raise1.5ex\hbox{$\leftrightarrow$}\mkern-16.5mu #1}

\def\half{{\textstyle{1\over2}}} 
\def\roughly#1{\raise.3ex\hbox{$#1$\kern-.75em\lower1ex\hbox{$\sim$}}}

\lref\berksimp{
N.~Berkovits,
``Simplifying and Extending the AdS(5) x S**5 Pure Spinor Formalism,''
JHEP {\bf 09}, 051 (2009)
[arXiv:0812.5074 [hep-th]].}

\lref\berkpert{
N.~Berkovits,
``Perturbative Super-Yang-Mills from the Topological AdS(5) x S**5 Sigma Model,''
JHEP {\bf 09}, 088 (2008)
[arXiv:0806.1960 [hep-th]].}

\lref\berkvafa{
N.~Berkovits and C.~Vafa,
``Towards a Worldsheet Derivation of the Maldacena Conjecture,''
JHEP {\bf 03}, 031 (2008)
[arXiv:0711.1799 [hep-th]].}

\lref\berklimit{
N.~Berkovits,
``A New Limit of the AdS(5) x S**5 Sigma Model,''
JHEP {\bf 08}, 011 (2007)
[arXiv:hep-th/0703282 [hep-th]].}

\lref\berktwistor{
N.~Berkovits,
``An Alternative string theory in twistor space for N=4 superYang-Mills,''
Phys. Rev. Lett. {\bf 93}, 011601 (2004)
[arXiv:hep-th/0402045 [hep-th]].}

\lref\berkads{
N.~Berkovits and O.~Chandia,
``Superstring vertex operators in an AdS(5) x S**5 background,''
Nucl. Phys. B {\bf 596}, 185-196 (2001)
[arXiv:hep-th/0009168 [hep-th]].}

\lref\berkderive{
N.~Berkovits,
``Sketching a Proof of the Maldacena Conjecture at Small Radius,''
JHEP {\bf 06} (2019), 111
[arXiv:1903.08264 [hep-th]].
}

\lref\gopa{
M.~R.~Gaberdiel and R.~Gopakumar,
``String Dual to Free N=4 Supersymmetric Yang-Mills Theory,''
Phys. Rev. Lett. {\bf 127}, no.13, 131601 (2021)
[arXiv:2104.08263 [hep-th]]\semi
M.~R.~Gaberdiel and R.~Gopakumar,
``The worldsheet dual of free super Yang-Mills in 4D,''
JHEP {\bf 11}, 129 (2021)
[arXiv:2105.10496 [hep-th]].}

\lref\gopatwo{
R.~Gopakumar and E.~A.~Mazenc,
``Deriving the Simplest Gauge-String Duality -- I: Open-Closed-Open Triality,''
[arXiv:2212.05999 [hep-th]].
}

\lref\gopavafa{
R.~Gopakumar and C.~Vafa,
``On the gauge theory / geometry correspondence,''
Adv. Theor. Math. Phys. {\bf 3}, 1415-1443 (1999)
[arXiv:hep-th/9811131 [hep-th]].}

\lref\vafa{
H.~Ooguri and C.~Vafa,
``World sheet derivation of a large N duality,''
Nucl. Phys. B {\bf 641}, 3-34 (2002)
[arXiv:hep-th/0205297 [hep-th]].}

\lref\gaiotto{
D.~Gaiotto and L.~Rastelli,
``A Paradigm of open / closed duality: Liouville D-branes and the Kontsevich model,''
JHEP {\bf 07}, 053 (2005)
[arXiv:hep-th/0312196 [hep-th]].}

\lref\grone{
D.~Gaiotto, N.~Itzhaki and L.~Rastelli,
``Closed strings as imaginary D-branes,''
Nucl. Phys. B {\bf 688}, 70-100 (2004)
[arXiv:hep-th/0304192 [hep-th]].}

\lref\wittenb{
E.~Witten,
``Perturbative gauge theory as a string theory in twistor space,''
Commun. Math. Phys. {\bf 252}, 189-258 (2004)
[arXiv:hep-th/0312171 [hep-th]].}

\lref\vafaa{
A.~Neitzke and C.~Vafa,
``N=2 strings and the twistorial Calabi-Yau,''
[arXiv:hep-th/0402128 [hep-th]].}

\lref\howe{
P.~S.~Howe and P.~C.~West,
``Nonperturbative Green's functions in theories with extended superconformal symmetry,''
Int. J. Mod. Phys. A {\bf 14} (1999), 2659-2674
[arXiv:hep-th/9509140 [hep-th]].}

\lref\wittop{
E.~Witten,
``Topological Sigma Models,''
Commun. Math. Phys. {\bf 118}, 411 (1988).}

\lref\phases{
E.~Witten,
``Phases of N=2 theories in two-dimensions,''
Nucl. Phys. B {\bf 403}, 159-222 (1993)
[arXiv:hep-th/9301042 [hep-th]].}

\lref\aisaka{
Y.~Aisaka and Y.~Kazama,
``Operator mapping between RNS and extended pure spinor formalisms for superstring,''
JHEP {\bf 08} (2003), 047
[arXiv:hep-th/0305221 [hep-th]].}

\lref\topmalda{
N.~Berkovits,
``Topological A-Model for $AdS_5\times S^5$ Superstring and the Maldacena Conjecture,''
[arXiv:2506.10907 [hep-th]].}

\lref\knighton{B.~Knighton,
``Classical geometry from the tensionless string,''
JHEP {\bf 05}, 005 (2023)
doi:10.1007/JHEP05(2023)005
[arXiv:2207.01293 [hep-th]].}

\lref\feynman{
R.~Gopakumar, R.~Kaushik, S.~Komatsu, E.~A.~Mazenc and D.~Sarkar,
``Strings from Feynman Diagrams,''
[arXiv:2412.13397 [hep-th]].}

\lref\gopathree{
R.~Gopakumar,
``From free fields to AdS: III,''
Phys. Rev. D {\bf 72}, 066008 (2005)
doi:10.1103/PhysRevD.72.066008
[arXiv:hep-th/0504229 [hep-th]].}

\lref\kontsevich{
M.~Kontsevich,
``Intersection theory on the moduli space of curves and the matrix Airy function,''
Commun. Math. Phys. {\bf 147}, 1-23 (1992).}

\lref\mukhi{
S.~Mukhi,
``Topological matrix models, Liouville matrix model and c = 1 string theory,''
[arXiv:hep-th/0310287 [hep-th]].}

\lref\motl{
N.~Berkovits and L.~Motl,
``Cubic twistorial string field theory,''
JHEP {\bf 04}, 056 (2004)
doi:10.1088/1126-6708/2004/04/056
[arXiv:hep-th/0403187 [hep-th]].}

\def\a{{\alpha}}

\def\ad{{\dot a}}

\def\Ab{{\overline A}}

\def\kb{{\bar \kappa}}

\def\d{{\delta}}
\def\e{{\epsilon}}

\def\k{{\kappa}}
\def\kb{{\overline\kappa}}

\def\Ab{{\overline A}}

\def\zb{{\overline z}}

\def\Ib{{\overline I}}

\def\L{{\Lambda}}
\def\Lb{{\overline\Lambda}}
\def\rhob{{\overline\rho}}

\def\half{{1\over 2}}
\def\p{{\partial}}

\def\pb{{\overline\partial}}

\def\ad{{\dot \a}}

\def\S{{\Sigma}}

\def\jb{{\overline j}}

\def\Phib{{\overline \Phi}}

\def\Lb{{\overline\Lambda}}
\def\rhob{{\overline\rho}}

\Title{\vbox{\baselineskip12pt
\hbox{}}}
{{\vbox{\centerline{Instanton Solutions of the Topological }
\smallskip
\centerline{A-model Dual to Free Super-Yang-Mills}}} }
\bigskip\centerline{Nathan Berkovits\foot{e-mail: nathan.berkovits@unesp.br}}
\bigskip
\centerline{\it ICTP South American Institute for Fundamental Research}
\centerline{\it Instituto de F\'\i sica Te\'orica, UNESP - Univ. 
Estadual Paulista }
\centerline{\it Rua Dr. Bento T. Ferraz 271, 01140-070, S\~ao Paulo, SP, Brasil}
\bigskip

\vskip .1in

A topological A-model constructed from supertwistor variables and a worldsheet U(1) gauge field was recently proposed to describe the $AdS_5\times S^5$ superstring at zero radius which is dual to free ${\cal N}$=4 d=4 super-Yang-Mills. In this note, holomorphic worldsheet instanton solutions of the supertwistor variables are constructed where the U(1) worldsheet gauge field is identified with the square-root of the Strebel differential that describes the ribbon graph on the worldsheet.

\vskip .1in

\Date {February 2026}
\newsec{Introduction}

In \topmalda, a topological A-model with N=(2,2) worldsheet supersymmetry was proposed to describe the $AdS_5\times S^5$ superstring at zero radius which is dual to free super-Yang-Mills. The worldsheet action for this topological A-model is
\eqn\action{S = \int d^2 z \int d^2 \k^+ \int d^2 \k^- (\Phib_\S ~e^V ~\Phi^\S )}
where $V(z, \zb;\k^\pm, \kb^\pm)$ is the real N=(2,2) worldsheet superfield for a U(1) gauge multiplet, 
$\Phi^\S$ and $\Phib_\S$ are chiral and antichiral N=(2,2) worldsheet superfields of U(1) charge $+1$ and $-1$ in the fundamental and anti-fundamental representation of $PSU(2,2|4)$, and $\S = (\a, \ad, j, \jb)$ is a $U(2,2|4)$ index with $(\a, \ad) = 1$ to 2 describing the $SU(2,2)$ components and $(j, \jb)=1$ to 2 describing the $SU(4)$ components. 

The equation of motion from varying the vector multiplet is $\Phib_\S \Phi^\S=0$, which can be solved by choosing $\Phi^\ad=\Phi^\jb = \Phib_\a = \Phib_j=0$. This solution spontaneously breaks the $PSU(2,2|4)$ symmetry to $PSU(1,1|2)\times PSU(1,1|2)$, and the remaining
superfields $\Phi^I =(\Phi^\a, \Phi^j)$ and $\Phib_\Ib= (\Phi_\ad, \Phi_\jb)$ transform linearly under the $PSU(1,1|2)\times PSU(1,1|2)$ subgroup and were used in \topmalda\ to construct the closed superstring vertex operators dual to traces of free super-Yang-Mills fields. 

For a topological A-model, states in the BRST cohomology will only depend on the $\k^\pm=\kb^\pm=0$ components of $\Phi^I$ and $\Phib_\Ib$ called $\L^I$ and $\Lb_\Ib$, which must satisfy the
holomorphic and antiholomorphic instanton equations
\eqn\insteq{ (\pb - i \Ab) \L^I = (\p +i A) \Lb_\Ib=0}
where $(A, \Ab)$ is the U(1) gauge field in $V$. As shown by Gaberdiel and Gopakumar in \gopa, the closed string states dual to traces of $k$ super-Yang-Mills fields can be described by $k$ modes of the $(\L^I, \Lb_\Ib)$ variables. And in the A-model proposal of \topmalda, these $k$ modes came from instanton solutions where the closed string vertex operator for the state contains ${k\over{2}}$ units of U(1) flux (or ``vortex charge"). For topological amplitudes with $N$ closed string vertex operators, the total U(1) flux on the surface is $K =\half \sum_{r=1}^N k_r$ and integration over $K$ instanton modes of $(\L^I, \Lb_\Ib)$ produces the $K$ propagators in the corresponding super-Yang-Mills Feynman diagram.

In this note, the instanton solutions containing these modes of $(\L^I, \Lb_\Ib)$ will be constructed using a U(1) gauge field $(A,\Ab)$ identified with the square-root of the Strebel differential on the worldsheet. The Strebel differential plays an important role in open-closed dualities by relating the closed string worldsheet with the open string ribbon graphs, e.g.  \kontsevich\mukhi\gopathree\gopatwo\feynman. The existence of instanton soluions for $(\L^I, \Lb_\Ib)$ where $(\L^I, \Lb_\Ib)$ are single-valued at the vertices will imply that only ``arithmetic" Riemann surfaces contribute to the topological closed string amplitudes that compute the free super-Yang-Mills correlation functions. Similar restrictions to ``arithmetic" Riemann surfaces were discussed in the open-closed dualities of \feynman, and although the square-root of the Strebel differential does not appear to have been previously related to a U(1) gauge field, \knighton\ describes supertwistor variables for $AdS_3\times S^3$ which satisfy a similar equation to \insteq. 

\newsec{Worldsheet gauge field}

To compute the $N$-point super-Yang-Mills correlation function $\langle \prod_{r=1}^N Tr ( Z_1^{(r)} ... Z_{k_r}^{(r)})\rangle$ where 
$Tr ( Z_1^{(r)} ... Z_{k_r}^{(r)})$ is a gauge-invariant state constructed from $k_r$ free super-Yang-Mills fields $Z_i^{(r)}$, the corresponding Riemann surface is obtained by gluing together
$N$ faces where the $r^{\rm th}$ face has $k_r$ edges of equal length. The center of the $r^{\rm th}$ face will be labeled by the point $z_r$, and the U(1) gauge field will contain flux $F$
concentrated at these points $z_r$ with the value ${1\over{2 \pi}}F(z,\zb) =\half \sum_{r=1}^N  k_r \d^2 (z-z_r)$. Note that this gluing produces a diagram on the Riemann surface which is dual to the associated super-Yang-Mills Feynman diagram, i.e. the vertices of the Feynman diagram are replaced by faces and the faces of the Feynman diagram are replaced by vertices.

Since the total flux ${1\over{2\pi}}\int d^2 z F = \half \sum_{r=1}^N k_r$ is non-zero, the U(1) gauge field $(A,\Ab)$ cannot be single-valued on the Riemann surface. However, it will now be shown that one
can choose ${4\over{\pi^2}} A^2$ to be single-valued and equal to the meromorphic Strebel differential where $A$ flips sign when crossing any edge from one face to an adjoining face. In other words, if the Riemann surface is composed of $N$ patches containing the $N$ faces, the gauge fields $A_{(1)}$ and $A_{(2)}$ on the overlap of patch $(1)$ and patch $(2)$ are related by $A_{(1)} = -A_{(2)}$. 

The Strebel differential 
$s(z) dz^2$ is defined to have double poles near $z=z_r$ of the form $s(z) \to -({{k_r}\over {2\pi}})^2 (z-z_r)^{-2}$ for $r=1$ to $N$. So on the $r^{\rm th}$ face near $z=z_r$, $A\to -i {{k_r}\over 4} (z-z_r)^{-1}$ and $\Ab \to i {{k_r}\over4} (\zb - \zb_r)^{-1}$ so that ${1\over{2\pi}} F = {i\over{2\pi}}(\pb A - \p \Ab) = \half k_r\delta^2 (z-z_r)$. 
If one defines the edges of the faces to correspond to the ``critical horizontal trajectories" that connect the zeros of $s(z)$, then $\sqrt{s(z)}$ is real and positive on these trajectories so that $A=\Ab$ on all edges. Defining the length of each edge by $\int_E  \sqrt{s} =  {1\over{\pi}}\int_E (A + \Ab) ={2\over\pi}\int_E A $ where $\int_E$ is an integral in the clockwise direction along the edge, the requirement that each of the $k_r$ edges on the $r^{\rm th}$ face has equal length implies that each edge has length $1$. Since the critical horizontal trajectories connecting the zeros of $s(z)$ include an integer number of edges, the distance between zeros of $s(z)$ is always an integer number which is the definition of an ``arithmetic" Riemann surface. 

It is easy to see that $A$ and $\Ab$ must flip sign when crossing an edge adjoining two faces since going in the clockwise direction along the edge on one face corresponds to going in the anticlockwise direction on the adjoining face. As an example, consider the case where one has six faces which each have four edges and are glued together to form a cube. The
Strebel differential will have the form 
\eqn\cube{ s(z) = c {{(z-y_1) ... (z-y_8)}\over {(z-z_1)^2 ... (z-z_6)^2}}}
where $(z_1, ..., z_6)$ label the centers of each face, $(y_1, ..., y_8)$ label the corners of the cube, and $c$ is some constant. So 
\eqn\Achoice{A(z) = \pm {{2\sqrt c}\over\pi} {{(z-y_1)^\half ... (z-y_8)^\half}\over {(z-z_1) ... (z-z_6)}}. }
Choosing the branch cuts of the square-roots in \Achoice\ to be along the 4 vertical edges of the cube, one finds that $A(z)$ will flip sign when crossing any edge if one chooses the $\pm$ in \Achoice\ to be the plus sign for the top and bottom faces of the cube and to be the minus sign for the other four faces of the cube.

\newsec{Holomorphic instanton solutions}

With $A(z)$ and $\Ab(\zb)$ determined, one can explicitly construct solutions to \insteq\ by defining
\eqn\definst{ \L^I (z, \zb)= \exp (i \int^z A + i\int^\zb \Ab) ~\rho^I(z), \quad \Lb_\Ib = \exp (-i \int^z A - i\int^\zb \Ab) ~\rhob_\Ib (\zb)}
where 
$(\rho^I(z), \rhob_\Ib(\zb))$ is required to be single-valued when $z$ crosses an edge. 
Since $(A, \Ab)$ flips sign when crossing an edge, $\L^I \to e^{-2 i (\int^z A + \int^\zb \Ab)} \L^I$ when crossing an edge. This is the desired gauge transformation for a variable of $+1$ U(1) charge where the gauge parameter on the edge is $\xi =  2 (\int^z A + \int^\zb \Ab) $ which satisfies $(A - \p\xi, \Ab - \pb \xi) = (-A, -\Ab)$. Furthermore, since all edges have integer length on an arithmetic Riemann surface, $\exp (2i \int^z A + 2i\int^\zb \Ab)$ is single-valued when $(z,\zb)$ is taken around any non-trivial cycle and $\exp(2i\int_E (A+\Ab))=1$ when the integral is along any edge $E$. This implies that on an arithmetic surface,
$(\L^I, \Lb_I)$ can be chosen to be single-valued at all vertices where edges meet. In other words, if the gauge parameter $\xi$ is chosen to vanish at one of the vertices, then $\exp(2i\int_E (A+\Ab))=1$ implies that $e^{i\xi}=1$ at all of the vertices.

Near $z=z_r$, $\Ab \to i {{k_r}\over 4}(\zb - \zb_r)^{-1}$, or more precisely,
$\Ab \to \lim_{\e\to 0} i{{ k_r}\over 4} {{z-z_r}\over{ |z-z_r|^2 + \e^2}}$. 
So \insteq\ implies that $\L^I \to \lim_{\e\to 0} f^I(z) (|z-z_r|^2 + \e^2)^{-{{k_r}\over 4}}$ near $z=z_r$ where $f^I$ is holomorphic. And since
$ \exp (i \int^z A + i \int^\zb \Ab) \to (z - z_r)^{{{k_r}\over 4}}(\zb - \zb_r)^{-{{k_r}\over 4}}$ near $z=z_r$, $\rho^I(z)$ can have poles of order $(z-z_r)^{-{{ k_r}\over 2}}$. The most general solution of \insteq\ on a genus zero surface is therefore
\eqn\solin{ \rho^I (z)= \prod_{r=1}^N (z-z_r)^{-\half k_r} ~ (\rho^I_{(0)} + z \rho^I_{(1)} + z^2 \rho^I_{(2)} + ... + z^K \rho^I_{(K)})}
where $K = \half \sum_{r=1}^N k_r ={1\over{2\pi}} \int d^2 z F$ is the total U(1) flux on the surface and $\rho^I (z)$ must be regular when $z\to \infty$.
This solution at genus zero can alternatively be parameterized through the parameters  $\rho_{(t)}^I = \rho^I(w_t)$ for $t=0$ to $K$ as 
\eqn\solina{ \rho^I (z)= \prod_{r=1}^N (z-z_r)^{-\half k_r}  \prod_{s=0}^K (z - w_s) \sum_{t=0}^K   {{\rho_{(t)}^I }\over { (z-w_t) f_t}}}
where $w_s$ is the midpoint of edge $s$ for $s=1$ to $K$, $w_0$ is an arbitrary fixed point on the worldsheet, and $f_t\equiv 
\prod_{r=1}^N (w_t-z_r)^{-\half k_r}  \prod_{s\neq t} (w_t - w_s)$.

At genus $g$, this solution generalizes to 
\eqn\soling{ \rho^I (z)=
 \prod_{r=1}^N E(z,z_r)^{-\half k_r} \prod_{s=0}^K E(z ,w_s) 
\sum_{t=0}^K   {{\rho_{(t)}^I }\over {E(z, w_t) f_t}}}
$$ \exp( \sum_{J,K=1}^g i (Re ~h_{(t)J}) (Im~ \tau)^{-1}_{JK} \int^z \omega_K  )$$
where $\omega_K$ are the $g$ holomorphic one-forms, $\tau_{JK}$ is the period matrix, $E(z_1, z_2)$ is the holomorphic prime form which transforms as $E(z_1, z_2) \to \exp (-i\pi \tau_{JJ} + 2\pi i \int^{z_1}_{z_2}\omega_J) E(z_1, z_2)$ when $z_1$ goes around the $J^{th}$ B-cycle of the genus $g$ surface, and 
\eqn\htt{h_{(t)J} = \half \sum_{r=1}^N k_r\int_z^{z_r} \omega_J -\sum_{s\neq t} \int_z^{w_s} \omega_J.}
The last line in \soling\ has been included so that the instanton parameters $(\rho_{(t)}^I, \rhob_{(t)\Ib})$ transform by a phase as 
\eqn\gaugee{\rho_{(t)}^I \to \exp(i \alpha_t) \rho_{(t)}^I, \quad \rhob_{(t) \Ib} \to \exp(-i \alpha_t) \rhob_{(t)\Ib}}
when $z$ goes around either an $A$-cycle or $B$-cycle in \soling. For example, when $z$ goes around the $J^{th}$ $B$-cycle, the first line in \soling\ picks up a factor of $\exp( h_{(t)J})$ which is independent of $z$. And because of the transformation $\int^z \omega_K \to \int^z \omega_K + 
\tau_{JK}$ in the second line in \soling, one finds $\alpha_t = Im~ h_{(t)J} +\sum_{K,L=1}^g (Re~h_{(t)K}) (Im~\tau)^{-1}_{KL}(Re~\tau)_{LJ}$ in \gaugee. Moreover, if $z$ goes around the $J^{th}$ $A$-cycle, $\int^z \omega_K \to \int^z \omega_K + \delta_{JK}$ implies that 
$\alpha_t = \sum_{K=1}^g (Re~h_{(t)K})(Im~ \tau)^{-1}_{JK} $ in \gaugee.

Since U(1) gauge invariance implies that all operators are invariant under \gaugee, the instanton solution of \soling\ is well-defined. 
So at any genus, the holomorphic and antiholomorphic instanton solutions of \insteq\ are described by $4K+4$ bosonic parameters and $4K+4$ fermionic parameters. As explained in \topmalda, integration over the instanton parameters 
$(\rho_{(t)}^I, \rhob_{(t)\Ib})$ for $t=1$ to $K$ reproduce the $K$ propagators in the corresponding super-Yang-Mills Feynman diagram. And since nothing depends on $(\rho_{(0)}^I, \rhob_{(0)\Ib})$,  integration over the remaining 4 bosonic and 4 fermionic instanton parameters $(\rho_{(0)}^I, \rhob_{(0)\Ib})$ gives a factor of $(0 \times \infty)^4$ which can be regularized to 1. This last integration in twistor space is reminiscent of integration over the constant modes of instantons in ordinary spacetime which decouple if the theory is translation invariant.

\vskip 10pt
{\bf Acknowledgements:}
I would like to thank Matthias Gaberdiel and Rajesh Gopakumar for useful discussions, and 
CNPq grant 311434/2020-7
and FAPESP grants 2016/01343-7, 2021/14335-0, 2019/21281-4 and 2019/24277-8 for partial financial support.

\footatend\vfill\supereject\immediate\closeout\rfile\writestoppt
\baselineskip=14pt\centerline{{\bf References}}\bigskip{\frenchspacing%
\parindent=20pt\escapechar=` \input refs.tmp\vfill\eject}\nonfrenchspacing

\end
\bye